\documentclass[letterpaper, preprint, paper,11pt]{AAS}	% for preprint proceedings

\usepackage{bm}
\usepackage{amsmath}
\usepackage{subfigure}
\usepackage[colorlinks=true, pdfstartview=FitV, linkcolor=black, citecolor= black, urlcolor= black]{hyperref}
\usepackage{overcite}
\usepackage{footnpag}			      	% make footnote symbols restart on each page

\usepackage{yfonts}
\usepackage{amssymb}
\newcommand{\R}{\mathbb{R}}

\PaperNumber{20-469}

\begin{document}

\title{Using Products of Exponentials to Define (Draw) Orbits and More}

\author{Aryslan Malik\thanks{Visiting Assistant Professor, Department of Aerospace Engineering, Embry-Riddle Aeronautical University, Daytona Beach, FL, 32114},
Troy Henderson\thanks{Associate Professor, Department of Aerospace Engineering, Embry-Riddle Aeronautical University, Daytona Beach, FL, 32114}, and Richard Prazenica\thanks{Associate Professor, Department of Aerospace Engineering, Embry-Riddle Aeronautical University, Daytona Beach, FL, 32114}
}

\maketitle{}

\begin{abstract}
The Product of Exponentials (PoE) formula is a mathematical tool that is used extensively in robotics. The virtue of using the exponential mapping, Lie Algebra and screw theory is that it allows an elegant and concise way of describing the orientation and position of a body with respect to another body in a multi-body system. Although the PoE formula is mainly used in robotics, this work aims to demonstrate the utility of the PoE formula as an alternative method for defining and drawing orbits given an orbital elements set. The work also explores the first derivative of the adapted PoE formula in the framework of orbital mechanics, which allows obtaining the state of the satellite (position and velocity) from the orbital elements set using the developed formulation.
\end{abstract}

\section{Introduction}
\begin{figure}[h!] 
    \centering
    \includegraphics[scale=0.6]{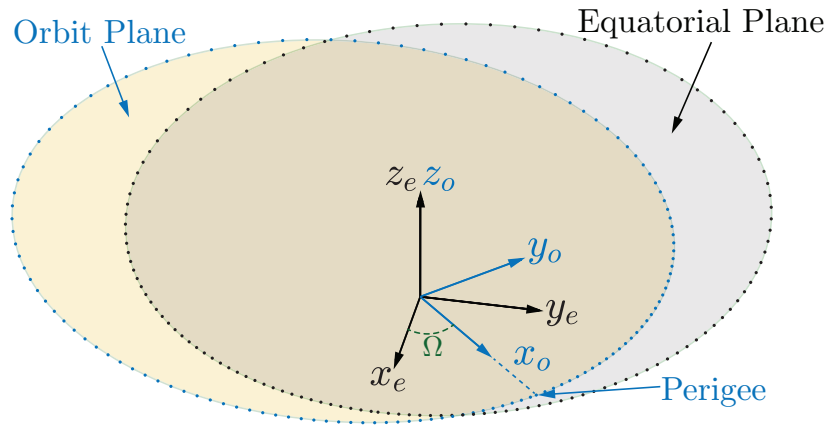}
    \caption{Longitude of Ascending Node}
    \label{fig:Omega}
\end{figure}

The PoE formulation was first described in Brockett's work, where it was used to describe forward kinematics of a multi-body system.\cite{POE1} Later, the computational aspects were discussed by Park.\cite{POE2} The underlying idea of the PoE formulation is to consider a joint as applying screw motion on the rest of the outward links (bodies).\cite{lynch2017modern,holm2009geometricmechanics,tsiotras1995new} To draw a parallel, in the case of a two-body system (\emph{e.g}. Earth and a satellite) the links are ``virtual" and the joints follow the orbital elements set $[\Omega, i, \omega, \theta, r(\theta)]$. Recent research focused on studying the error associated with Special Orthogonal $SO(n)$ transformations.\cite{attude_error} The celestial mechanics formulations and equations, solutions to the Keplerian two-body problem, are drawn from well established material.\cite{junkins2009analytical,beutler2004methods} To demonstrate the concept, first consider the inertial Earth frame to be $(x_e,y_e,z_e)$ and a satellite's frame to be $(x_o,y_o,z_o)$, which for now coincides with the inertial Earth frame. In the PoE formulation it can be described as follows:

\begin{equation}\label{explanation1}
{T_{eo}}=\begin{bmatrix}
           R_{eo} & p_{eo}\\
           0 & 1 \\
            \end{bmatrix}=\begin{bmatrix}
           1&0&0&0\\
           0&1&0&0\\
           0&0&1&0\\
           0&0&0&1\\ \end{bmatrix}=I_{4\times{}4}=M_{eo}
\end{equation}
where, $R_{eo}\in{SO(3)}$ is the orientation of the satellite represented as a member of the special orthogonal group, $p_{eo}\in\R^3$ is the position of the satellite represented as a $3\times{}1$ vector, and $T_{eo}\in{SE(3)}$ is the configuration of satellite represented as a member of the special euclidean group. The subscript $(eo)$ denotes that the orientation, position, and configuration of the satellite are given in the inertial Earth frame.
Since the satellite frame coincides with the inertial Earth frame when the orbital elements set is zero, the configuration given in Equation \ref{explanation1} is called as the ``home configuration" $M_{eo}$. Notice that, in the case of ``home configuration", the orientation $R_{eo}\in{SO(3)}$ is $I_{3\times{}3}$ identity matrix, which means that the satellite's orientation is the same as the inertial Earth frame's orientation, and position vector $p_{eo}\in\R^3$ is the $3\times1$ zero vector, which means that the origin of the satellite frame coincides with the inertial Earth frame origin. Now, for simplicity, let the satellite frame only have nonzero Longitude of Ascending Node, which would mean that the satellite frame is just the Earth frame rotated about the $z_{e}$-axis with $\Omega$ angle. This is demonstrated in Figure \ref{fig:Omega}, where the orientation of the satellite's orbit frame is rotated about the $z_{e}$-axis relative to the inertial Earth frame. The configuration of the satellite in the PoE formulation can be demonstrated as follows:
\begin{equation}\label{explanation1}
{T_{eo}}=e^{[\mathcal{S}_1]\Omega}M_{eo}
\end{equation}
\begin{equation}\label{eq9}
    \mathcal{S}_1 = \begin{bmatrix}
           \omega_1 \\
           v_1 \\
         \end{bmatrix}=\begin{bmatrix}
0 \\
0 \\
1 \\
0 \\
0 \\
0 \\
         \end{bmatrix}\in\R^6
\end{equation}
where, $\mathcal{S}_1\in\R^6$ is a ``unit" screw axis represented in the inertial Earth frame such that $\left\lVert \omega_{1}\right\rVert=1$ and $v_1$ is the linear velocity at the fixed-frame origin, expressed in the fixed frame produced purely due to the rotation about the screw axis $(v_{1}=-\omega_{1}\times{r_{q}})$, where $q$ is a point on the screw axis. In the case of $\mathcal{S}_1$, $r_{q}=0$ since the screw axis coincides with the inertial Earth frame. Square brackets represent the skew-symmetric mapping of an element such that: $p\in\R^3\,	{\rightarrow}\,{[p]}\in{so(3)}$ and ${\mathcal{S}}\in\R^6\,	{\rightarrow}\,{[\mathcal{S}]}\in{se(3)}$. 
\begin{align}\label{skew-sym}
    [\mathcal{S}]=\begin{bmatrix}
           [\omega] & v \\
           0 & 0 \\
         \end{bmatrix}\in{se(3)}
\end{align}
Thus, the exponential mapping of an element operates such that:  $[{\mathcal{S}}]\in{se(3)}\,	{\rightarrow}\,T\in{SE(3)}$.

If inclination $(i)$ and argument of periapsis $(\omega)$ are added to Equation \ref{explanation1}, the configuration of the satellite frame assumes the following form:
\begin{equation}\label{explanation3}
{T_{eo}}=e^{[\mathcal{S}_1]\Omega}e^{[\mathcal{S}_2]i}e^{[\mathcal{S}_3]\omega}M_{eo}
\end{equation}
where, 
\begin{equation}\label{eq9}
    \mathcal{S}_2 = \begin{bmatrix}
1         \\
0         \\
0         \\
0    \\
0         \\
0    \\
         \end{bmatrix}
         \in\R^6 \quad\mathcal{S}_3 = \begin{bmatrix}
0         \\
0         \\
1         \\
0    \\
0         \\
0    \\
         \end{bmatrix}\in\R^6
\end{equation}

\begin{figure}[h!] 
    \centering
    \includegraphics[scale=0.6]{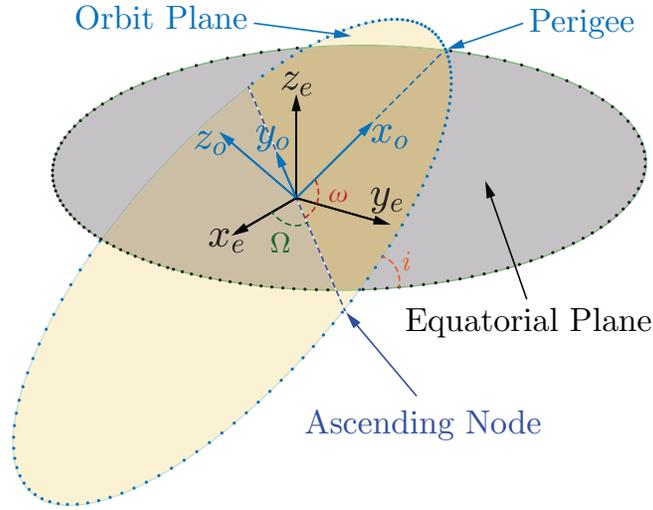}
    \caption{Perifocal frame of the satellite}
    \label{perifocal}
\end{figure}

Notice that the screw axes $\mathcal{S}_1, \mathcal{S}_2, \mathcal{S}_3$ actually correspond to the Euler $313$ sequence, which is also reflected on the screw axes vectors' elements. The configuration given in Equation \ref{explanation3} actually represents the Perifocal frame of the satellite. This is demonstrated in Figure \ref{perifocal}.

In the next section, the development of the PoE formulation in the orbital mechanics framework is continued with an example scenario. Then, the velocity representation using the PoE formulation is developed, and
some numerical test scenarios are introduced. Finally, conclusions are derived based on the
results provided by the comparison tests.

\section{General Formulation}
Essentially the idea of the PoE formulation is that the two-body system acts like a ``robotic arm" with joints that correspond to the orbital elements set: $[\Omega, i, \omega, \theta, r(\theta)]$. So, the first four elements act like single degree of freedom revolute joints, and the last $r(\theta)$ acts like a prismatic joint. The beauty of this method is that the configuration (orientation and position) of a satellite is given by only one equation. The configuration of a satellite given in the inertial Earth frame is as follows:
\begin{equation}\label{eq:bodyframe}
{T_{eo}}=e^{[\mathcal{S}_1]\Omega}e^{[\mathcal{S}_2]i}e^{[\mathcal{S}_3]\omega}e^{[\mathcal{S}_4]\theta}e^{[\mathcal{S}_5]r(\theta)}M_{eo}
\end{equation}
where, 
\begin{equation}\label{rtheta}
r(\theta)=\frac{a(1-e^2)}{1+e\cos\theta}
\end{equation}
\begin{equation}
M_{eo}=\begin{bmatrix}
           1&0&0&0\\
           0&1&0&0\\
           0&0&1&0\\
           0&0&0&1\\ \end{bmatrix}=I_{4\times{}4}
\end{equation}

\begin{align}
    \mathcal{S}_1 =\begin{bmatrix}
0 \\
0 \\
1 \\
0 \\
0 \\
0 \\
         \end{bmatrix}
    \mathcal{S}_2 = \begin{bmatrix}
1         \\
0         \\
0         \\
0    \\
0         \\
0    \\
         \end{bmatrix}
    \mathcal{S}_3 =& \begin{bmatrix}
0 \\
0 \\
1 \\
0 \\
0 \\
0 \\
         \end{bmatrix}\\
    \mathcal{S}_4 = \begin{bmatrix}
0 \\
0 \\
1 \\
0 \\
0 \\
0 \\
         \end{bmatrix}
    \mathcal{S}_5 =& \begin{bmatrix}
0       \\
0       \\
0       \\
1       \\
0   \\
0  \\
         \end{bmatrix}\nonumber
         \end{align}
where $a$ is the semi-major axis and $e$ is the eccentricity. The subscript $(eo)$ denotes that $M_{eo}$ is the ``home" configuration of an ``object" relative to Earth. $M_{eo}$ is the $4\times{}4$ identity matrix because initially it is assumed that the object's position and orientation coincides with the Earth's position and orientation. 
\begin{figure}[h!] 
    \centering
    \includegraphics[scale=0.6]{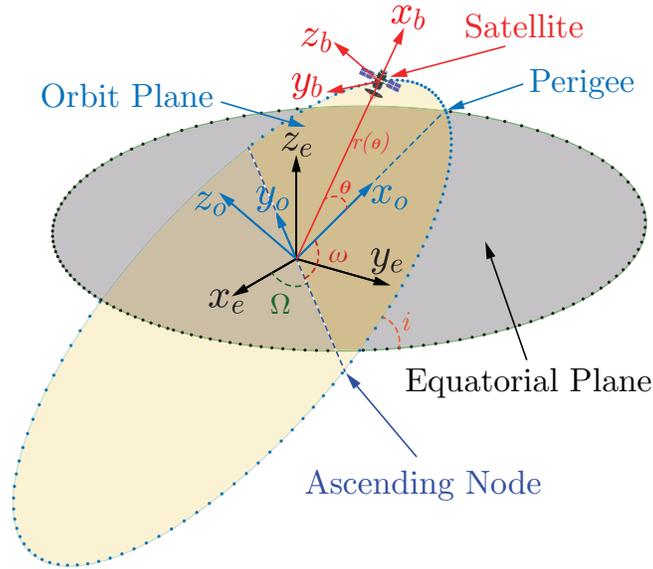}
    \caption{Body frame of the satellite}
    \label{bodyframe}
\end{figure}

\begin{table}[htbp]
	\fontsize{10}{10}\selectfont
    \caption{Coordinate Systems}
   \label{tab:Coordinate_systems}
        \centering 
   \begin{tabular}{l | l} % Column formatting, 
      \hline 
       Coordinate system & Configuration\\
      \hline 
 Inertial (Earth)$(x_e, y_e, z_e)$& $T=M_{eo}$\\ 
 Orbit           $(x_o, y_o, z_o)$&${T}=e^{[\mathcal{S}_1]\Omega}e^{[\mathcal{S}_2]i}e^{[\mathcal{S}_3]\omega}M_{eo}$\\
 Body            $(x_b, y_b, z_b)$&${T}=e^{[\mathcal{S}_1]\Omega}e^{[\mathcal{S}_2]i}e^{[\mathcal{S}_3]\omega}e^{[\mathcal{S}_4]\theta}e^{[\mathcal{S}_5]r(\theta)}M_{eo}$\\
      \hline
   \end{tabular}
\end{table}

Generally speaking, Equation \ref{explanation3} will result in the orbital frame $(x_{o},y_{o},z_{o})$, and Equation \ref{eq:bodyframe} in the body frame $(x_{b},y_{b},z_{b})$ as shown in Figure \ref{bodyframe} and Table \ref{tab:Coordinate_systems}.

\begin{figure}[h!]
    \centering
    \includegraphics[scale=0.9]{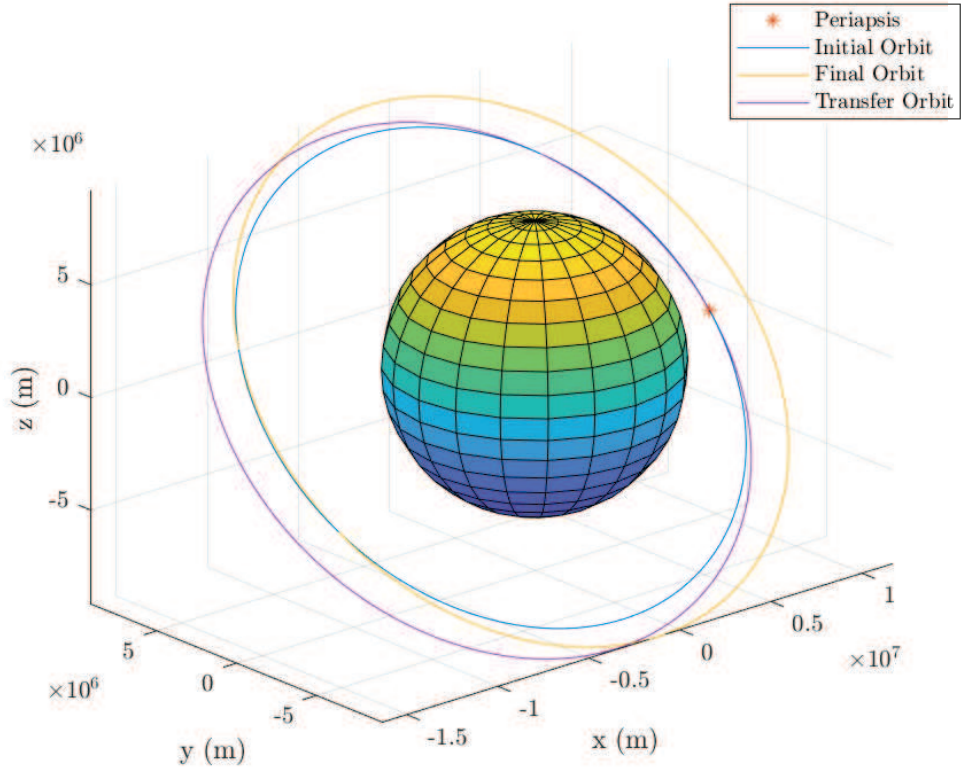}
    \caption{Transfer Orbit}
    \label{fig:1}
\end{figure}

To test the PoE formulation, three orbits were plotted: an initial orbit, a final orbit, and a transfer orbit that must occur at $120$ degrees of true anomaly $(\theta)$. The orbits are described in Table \ref{tab:label}. The plot of the transfer orbit with 45 degrees of inclination $(i)$, and roughly 6 degrees of argument of periapsis $(\omega)$, along with initial and final orbits, is shown in Figure \ref{fig:1}.   

\begin{table}[htbp]
	\fontsize{10}{10}\selectfont
    \caption{Description of Orbits}
   \label{tab:label}
        \centering 
   \begin{tabular}{c | r | r | r} % Column formatting, 
      \hline 
       Orbital Element & Initial Orbit & Final Orbit &  Transfer Orbit\\
      \hline 
 $a$      & $12000$ km   & $13000$ km    & 12993 km\\ 
 $e$      & $0.2$        & $0.1$         & $0.262$\\
 $i$      & $45^{\circ}$ & $45^{\circ}$  & $45^{\circ}$\\
 $\omega$ & $0^{\circ}$  & $0^{\circ}$   & $5.96^{\circ}$\\
 $\Omega$ & $0^{\circ}$  & $0^{\circ}$   & $0^{\circ}$\\
 $\theta$ & $0^{\circ}$  & $120^{\circ}$ & N/A\\
      \hline
   \end{tabular}
\end{table}

The results of the PoE formulation were checked by comparing the $(x,y,z)$ components of the $r(\theta)$ vector with the ones obtained using integration (classical computations). Also, by setting inclination to zero, the results of the PoE formulation were compared with the straightforward $r(\theta)$ computation given in Equation \ref{rtheta}. All methods gave the same orbits and components of the radius vector and its magnitude $r(\theta)$.

\section*{Development of Velocity Representation Using PoE Formulation}
Interestingly, by using the PoE formulation, the velocity of an object or satellite can be found by knowing the rate of change of true anomaly $(\Dot{\theta})$. The inertial velocity represented as a spatial twist is given by:
\begin{equation}\label{vs1}
    \mathcal{V}_s=\begin{bmatrix}
           \omega_s\\
           v_s\\
           \end{bmatrix}\in\R^6
\end{equation}
The last three elements of this vector represent the inertial velocity of the satellite not taking into account the rotation of the body frame relative to the inertial frame, and the first three represent angular velocity components expressed in the inertial frame. Thus, the velocity vector of the satellite expressed in the inertial frame is obtained from the spatial twist via the following formula:
\begin{equation}\label{inertial_velocity}
    \textgoth{V}_s=v_s+\omega_s\times{p_{eo}}\in\R^3
\end{equation}

The $4\times4$ matrix representation of the $\mathcal{V}_s\in\R^6$ spatial twist vector is given as follows:
\begin{equation}\label{vs}
    [\mathcal{V}_s]=\begin{bmatrix}
           [\omega_s]&v_s\\
           0&0\\
           \end{bmatrix}=\dot{T}_{eo}T_{eo}^{-1}\in{se(3)}
\end{equation}
The body twist of the satellite can be represented in the body frame as follows:
\begin{equation}
    [\mathcal{V}_b]=T_{eo}^{-1}\dot{T}_{eo}=T_{eo}^{-1}[\mathcal{V}_s]T_{eo}
\end{equation}
where,
\begin{align}\label{8}\dot{T}_{eo}&=[\mathcal{S}_1]\dot{\Omega}e^{[\mathcal{S}_1]\Omega}e^{[\mathcal{S}_2]i}e^{[\mathcal{S}_3]\omega}e^{[\mathcal{S}_4]\theta}e^{[\mathcal{S}_5]r(\theta)}M_{eo}\nonumber\\&+e^{[\mathcal{S}_1]\Omega}[\mathcal{S}_2]\dot{i}e^{[\mathcal{S}_2]i}e^{[\mathcal{S}_3]\omega}e^{[\mathcal{S}_4]\theta}e^{[\mathcal{S}_5]r(\theta)}M_{eo}\nonumber\\&+e^{[\mathcal{S}_1]\Omega}e^{[\mathcal{S}_2]i}[\mathcal{S}_3]\dot{\omega}e^{[\mathcal{S}_3]\omega}e^{[\mathcal{S}_4]\theta}e^{[\mathcal{S}_5]r(\theta)}M_{eo}\\&+e^{[\mathcal{S}_1]\Omega}e^{[\mathcal{S}_2]i}e^{[\mathcal{S}_3]\omega}[\mathcal{S}_4]\dot{\theta}e^{[\mathcal{S}_4]\theta}e^{[\mathcal{S}_5]r(\theta)}M_{eo}\nonumber\\&+e^{[\mathcal{S}_1]\Omega}e^{[\mathcal{S}_2]i}e^{[\mathcal{S}_3]\omega}e^{[\mathcal{S}_4]\theta}[\mathcal{S}_5]\dot{r}(\theta)e^{[\mathcal{S}_5]r(\theta)}M_{eo}\nonumber
\end{align} 
\begin{equation}\label{9}
 T_{eo}^{-1}=M_{eo}^{-1}e^{-[\mathcal{S}_5]r(\theta)}e^{-[\mathcal{S}_4]\theta}e^{-[\mathcal{S}_3]\omega}e^{-[\mathcal{S}_2]i}e^{-[\mathcal{S}_1]\Omega}
\end{equation}
Combining Equations \ref{vs}, \ref{8} and \ref{9}:
\begin{align}\label{10}
 [\mathcal{V}_s]=[\mathcal{S}_1]\dot{\Omega}&+e^{[\mathcal{S}_1]\Omega}[\mathcal{S}_2]e^{-[\mathcal{S}_1]\Omega}\dot{i}+e^{[\mathcal{S}_1]\Omega}e^{[\mathcal{S}_2]i}[\mathcal{S}_3]e^{-[\mathcal{S}_2]i}e^{-[\mathcal{S}_1]\Omega}\dot{\omega}\nonumber\\&+e^{[\mathcal{S}_1]\Omega}e^{[\mathcal{S}_2]i}e^{[\mathcal{S}_3]\omega}[\mathcal{S}_4]e^{-[\mathcal{S}_3]\omega}e^{-[\mathcal{S}_2]i}e^{-[\mathcal{S}_1]\Omega}\dot{\theta}\\&+e^{[\mathcal{S}_1]\Omega}e^{[\mathcal{S}_2]i}e^{[\mathcal{S}_3]\omega}e^{[\mathcal{S}_4]\theta}[\mathcal{S}_5]e^{-[\mathcal{S}_4]\theta}e^{-[\mathcal{S}_3]\omega}e^{-[\mathcal{S}_2]i}e^{-[\mathcal{S}_1]\Omega}\dot{r}(\theta)\nonumber
\end{align}
Alternatively, Equation \ref{10} can be written in vector form as the following:
\begin{align}\label{11}
    \mathcal{V}_s&=\mathcal{S}_1\dot{\Omega}+\mathrm{Ad}_{e^{[\mathcal{S}_1]\Omega}}(\mathcal{S}_2)\dot{i}+\mathrm{Ad}_{e^{[\mathcal{S}_1]\Omega}e^{[\mathcal{S}_2]i}}(\mathcal{S}_3)\dot{\omega}\\&+\mathrm{Ad}_{e^{[\mathcal{S}_1]\Omega}e^{[\mathcal{S}_2]i}e^{[\mathcal{S}_3]\omega}}(\mathcal{S}_4)\dot{\theta}+\mathrm{Ad}_{e^{[\mathcal{S}_1]\Omega}e^{[\mathcal{S}_2]i}e^{[\mathcal{S}_3]\omega}e^{[\mathcal{S}_4]\theta}}(\mathcal{S}_5)\dot{r}(\theta)\nonumber
\end{align}

In the case where longitude of ascending node $\Omega$, inclination $i$, and argument of periapsis $\omega$ are constant for a given satellite's orbit, Equation \ref{11} simplifies further to the following:
\begin{equation}\label{velocity}
   \mathcal{V}_s=\mathrm{Ad}_{e^{[\mathcal{S}_1]\Omega}e^{[\mathcal{S}_2]i}e^{[\mathcal{S}_3]\omega}}(\mathcal{S}_4)\dot{\theta}+\mathrm{Ad}_{e^{[\mathcal{S}_1]\Omega}e^{[\mathcal{S}_2]i}e^{[\mathcal{S}_3]\omega}e^{[\mathcal{S}_4]\theta}}(\mathcal{S}_5)\dot{r}(\theta) 
\end{equation}
where,
\begin{equation}\label{rthetadot}
    \dot{r}(\theta) = \frac{dr(\theta)}{d\theta}\frac{d\theta}{dt}= \frac{dr(\theta)}{d\theta}\dot{\theta}=\frac{ae(1-e^2)\sin{\theta}}{(1+e\cos{\theta})^2}\dot{\theta}
\end{equation}
and, for any $T=(R,p)\in{SE(3)}$, the adjoint mapping is defined as
\begin{equation}\label{adjoint}
    \mathrm{Ad}_{T}=\begin{bmatrix}
           R&0\\
           [p]R&R\\
           \end{bmatrix}\in{\R^{6\times6}}
\end{equation}

Therefore, in order to obtain the inertial velocity of a satellite orbiting the Earth assuming that longitude of ascending node $\Omega$, inclination $i$, and argument of periapsis $\omega$ are constant, it is required to know the true anomaly $\theta$ and its rate of change $\dot{\theta}$ at that true anomaly. Provided the aforementioned information is known, the inertial velocity can be obtained using the adjoint mappings given in Equation \ref{velocity}. Assuming that there is no disturbance, such that the angular momentum vector remains constant, the rate of change of true anomaly can be obtained as follows:
\begin{equation}\label{heq}
    h=r^2\dot{\theta}
\end{equation}
\begin{equation}\label{heq2}
    h=\sqrt{\mu{}p}=\sqrt{\mu{}a(1-e^2)}
\end{equation}
where $\mu$ is the standard gravitational parameter for the Earth. Combining Equations \ref{rtheta}, \ref{heq} and \ref{heq2}:
\begin{equation}\label{thetadoteq}
    \dot{\theta}=\frac{\sqrt{\mu{}a(1-e^2)}}{r^2}=(1+e\cos\theta)^2\sqrt{\frac{\mu}{[a(1-e^2)]^3}}
\end{equation}
Thus, the time rate of change of the position $r(\theta)$, given in Equation \ref{rthetadot}, simplifies to the following:
\begin{equation}\label{rthetadot2}
    \dot{r}(\theta) =e\sin\theta\sqrt{\frac{\mu}{a(1-e^2)}}
\end{equation}

By combining Equations \ref{eq:bodyframe}, \ref{inertial_velocity}, \ref{velocity}, \ref{thetadoteq}, and \ref{rthetadot2}, it is possible to fully define the state of the satellite $(p_{eo},\textgoth{V}_{eo})$ in the inertial Earth frame or satellite body frame for any given value of true anomaly. Essentially, Equation \ref{velocity} gives the general formulation of spatial twist, which in turn allows the computation of the inertial velocity of a satellite as given in Equation \ref{inertial_velocity}. An additional virtue of using the general formulation is that it allows computation of the inertial velocity of a satellite for cases where all orbital elements vary with respect to time. However, in general, the longitude of ascending node, inclination, and argument of periapsis $(\Omega,i,\omega)$ are constant. Assuming that out of the orbital elements set only the true anomaly $\theta$ and the radius vector $r(\theta)$ change with time, the computation of the inertial velocity of the satellite can be simplified. Looking at Figure \ref{bodyframe}, as the motion happens in the orbital plane, the velocity components in the body frame are as follows:
\begin{equation}\label{velocitybody}
 \textgoth{V}_b=\begin{bmatrix}
           \dot{r}(\theta)\\
           r(\theta)\dot{\theta}\\
           0\\
           \end{bmatrix}
\end{equation}
Now, the velocity of the satellite expressed in the inertial frame can be shown to be as follows:
\begin{equation}\label{velocitybody}
 \textgoth{V}_s=R_{eo}\begin{bmatrix}
           \dot{r}(\theta)\\
           r(\theta)\dot{\theta}\\
           0\\
           \end{bmatrix}=R_{eo}\textgoth{V}_b
\end{equation}

\section{Conclusion}
This paper presents applications of the PoE formulation in the field of celestial mechanics. The proposed method shows that it is possible to obtain orbital states from the orbital elements set with a concise and elegant PoE formulation. With this approach, the position and orientation of an object is expressed as a $4\times4$ configuration matrix. The velocity of this object can be obtained from spatial twist (general formulation), or by a $3\times3$ transformation matrix in the case when, out of all orbital elements, only the true anomaly and the radius vector vary with time, which is generally the case. The state of the satellite or any object can be expressed either in the inertial or body frame. The reverse can be performed as well with the utility of well established methods.

%\section{Acknowledgment}

\bibliographystyle{AAS_publication}   % Number the references.
\bibliography{AAStemplatev2_0_6}   % Use references.bib to resolve the labels.

\end{document}